\makeatletter\def\graphicscache@inhibit{true}\makeatother

\documentclass{vgtc}                          %
\ifpdf%
  \pdfoutput=1\relax                   %
  \usepackage{graphicx}                %
  \DeclareGraphicsExtensions{.pdf,.png,.jpg,.jpeg} %
\else%
  \usepackage{graphicx}                %
  \DeclareGraphicsExtensions{.eps}     %
\fi%

\graphicspath{{figures/}{pictures/}{images/}{./}} %

\usepackage{microtype}                 %
\PassOptionsToPackage{warn}{textcomp}  %
\usepackage{textcomp}                  %
\usepackage{mathptmx}                  %
\usepackage{times}                     %
\usepackage{cite}                      %
\usepackage{tabu}                      %
\usepackage{booktabs}                  %

\usepackage{graphicscache}

\usepackage[bookmarks=false,hidelinks]{hyperref}

\usepackage{color}
\usepackage{comment}
\usepackage{subfigure}
\usepackage{amssymb}
\usepackage{multirow}

\usepackage{xcolor}
\usepackage{enumitem}
\usepackage{xspace}
\def\ie{i.e.\xspace}
\def\eg{e.g.\xspace}
\def\etal{et~al.\xspace}
\usepackage[normalem]{ulem}

\usepackage{tabularx}
\newcolumntype{L}[1]{>{\raggedright\arraybackslash}m{#1}}
\newcolumntype{C}[1]{>{\centering\arraybackslash}m{#1}}
\newcolumntype{R}[1]{>{\raggedleft\arraybackslash}m{#1}}

\usepackage{algorithm}
\usepackage{algpseudocode}
\algrenewcommand\algorithmicrequire{\textbf{Input:}}
\algrenewcommand\algorithmicensure{\textbf{Output:}}
\algnewcommand\And{\textbf{and}\xspace}
\algnewcommand\Or{\textbf{or}\xspace}
\algnewcommand\Not{\textbf{not}\xspace}

\newcommand{\abs}[1]{\ensuremath{\left\vert#1\right\vert}}

\usepackage{fixmath}
\newcommand{\makeMathBold}[1]{\mathbold{#1}}

\usepackage{tikz}

\onlineid{1293}

\vgtccategory{Research}

\vgtcinsertpkg

\title{Efficient 3D Reconstruction and Streaming for Group-Scale Multi-Client Live Telepresence}

\author{Patrick Stotko\thanks{e-mail: stotko@cs.uni-bonn.de} %
\and Stefan Krumpen\thanks{e-mail: krumpen@cs.uni-bonn.de} %
\and Michael Weinmann\thanks{e-mail: mw@cs.uni-bonn.de} %
\and Reinhard Klein\thanks{e-mail: rk@cs.uni-bonn.de}}
\affiliation{\scriptsize University of Bonn}

\teaser{
    \centering
    \includegraphics[width=\textwidth,height=0.3\textheight,keepaspectratio]{figures/teaser.pdf}
    \vspace{-3.3mm}
    \caption{Illustration of our novel highly scalable multi-client live telepresence system. While previous approaches are limited to a low number of up to 4 remote exploration clients, our system is capable of providing an immersive telepresence experience within a live-captured high-quality scene reconstruction to more than 24 clients simultaneously without introducing further latency.}
    \label{fig:teaser}
}

\abstract{
Sharing live telepresence experiences for teleconferencing or remote collaboration receives increasing interest with the recent progress in capturing and AR/VR technology.
Whereas impressive telepresence systems have been proposed on top of on-the-fly scene capture, data transmission and visualization, these systems are restricted to the immersion of single or up to a low number of users into the respective scenarios.
In this paper, we direct our attention on immersing significantly larger groups of people into live-captured scenes as required in education, entertainment or collaboration scenarios.
For this purpose, rather than abandoning previous approaches, we present a range of optimizations of the involved reconstruction and streaming components that allow the immersion of a group of more than 24 users within the same scene -- which is about a factor of 6 higher than in previous work -- without introducing further latency or changing the involved consumer hardware setup.
We demonstrate that our optimized system is capable of generating high-quality scene reconstructions as well as providing an immersive viewing experience to a large group of people within these live-captured scenes.
} %

\CCScatlist{
  \CCScatTwelve{Human-centered computing}{Human computer interaction (HCI)}{Interaction paradigms}{Virtual reality};
  \CCScatTwelve{Human-centered computing}{Human computer interaction (HCI)}{Interaction paradigms}{Collaborative interaction};
  \CCScatTwelve{Computing methodologies}{Computer graphics}{Graphics systems and interfaces}{Virtual reality};
  \CCScatTwelve{Computing methodologies}{Computer vision}{Computer vision problems}{Reconstruction}
}

\begin{document}

\firstsection{Introduction}

\maketitle

\begin{figure*}[t]
    \centering
    \hspace{0.1mm}
    \includegraphics[width=0.8\linewidth,height=0.3\textheight,keepaspectratio]{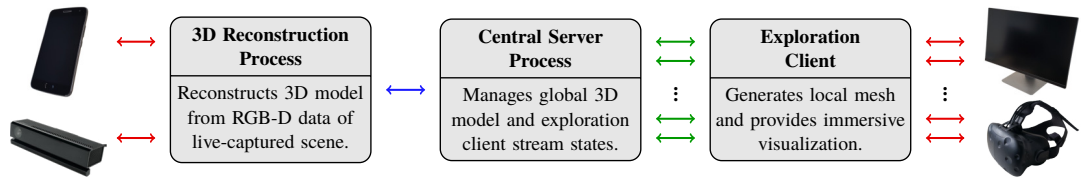}
    \caption{Overview of the major components of state-of-the-art live multi-client live telepresence systems. RGB-D image data acquired by a single camera device are streamed to a cloud server (red arrows) where a global 3D scene model is reconstructed in a dedicated 3D reconstruction process. This scene model is then passed to the central server process (blue arrows) which also runs on the cloud server and manages a bandwidth-optimized version of the model as well as the client states. Large groups of people (more than 24 in our system), each running an exploration client on their local hardware, can independently request parts of the reconstructed global scene model (green arrows) and render the locally generated mesh on their display devices (red arrows).}
    \label{fig:overview}
\end{figure*}

\begin{tikzpicture}[remember picture,overlay]%
\node[anchor=south,align=center,font=\mdseries,yshift=0.3cm, xshift=-0.3cm] at (current page.south) {%
\begin{minipage}[c]{\textwidth}\copyright \ 2019 IEEE. Personal use of this material is permitted. Permission from IEEE must be obtained for all other uses, in any current or future media, including reprinting/republishing this material for advertising or promotional purposes,creating new collective works, for resale or redistribution to servers or lists, or reuse of any copyrighted component of this work in other works. The final version of record is available at \url{http://dx.doi.org/10.1109/ISMAR.2019.00018}.\end{minipage}%
};%
\end{tikzpicture}%
The rapidly increasing potential of AR/VR technology has led to several highly advanced telepresence applications such as teleconferencing in room-scale environments~\cite{Orts-Escolano:2016,Fairchild:2016} or the exploration of places -- that may vary from the users' local physical environment -- for live-captured scenes beyond room-scale~\cite{stotko2019slamcast} and for remote collaboration purposes.
To meet the critical success factors of an immersive telepresence experience for on-the-fly captured 3D data as required by these scenarios, these systems impose strong demands regarding the reconstruction and streaming speed as well as the visual quality of the acquired scene.
Furthermore, the interactive exploration within the scene requires rendering at high framerates and low latency to avoid motion sickness.
This means that all of the involved processing steps including 3D scene capture, data transmission and visualization have to be achieved in real-time, while taking the typically available network bandwidth and client-side compute hardware into account.
Previous teleconferencing systems~\cite{Vasudevan:2011,Orts-Escolano:2016,Fairchild:2016} were designed to capture a fixed region of interest based on expensive well-calibrated acquisition setups involving statically mounted cameras.
In contrast, the live telepresence system by Stotko \etal~\cite{stotko2019slamcast} is tailored to the acquisition of scenes beyond such a fixed size defined by the setup and involves portable, consumer-grade capture hardware.
As a result, efficiently representing and transmitting the scene to visualization devices is significantly harder.
While further work has been spent on parallelized capturing~\cite{Golodetz2018Collaborative}, the goal of immersing a large number of people into the same live-captured environment while allowing them to interact with each other for \eg remote collaboration and exploration scenarios, to the best of our knowledge, has not received a lot of attention so far.
In particular, the major challenge is given by the accurate reconstruction and transmission of 3D models while keeping the computational burden as well as the memory and streaming requirements as low as possible, thus, minimizing the amount of unnecessary or unreliable model data resulting from noise and outliers in the captured input data.

In this paper, we address the scalability of live telepresence systems to the immersion of whole groups of (more than 24) people without introducing further latency as required for education, entertainment and collaboration scenarios (see \autoref{fig:teaser}).
For this purpose, rather than developing new techniques for 3D capture and data transmission, we demonstrate how existing well-established systems for 3D reconstruction and streaming can be optimized to significantly increase the scalability of live telepresence systems under strong constraints regarding latency and bandwidth.
We demonstrate that our extensions result in compact high-quality 3D reconstructions and finally allow the immersion of more than 24 people within the same live-captured scene (beyond room-scale), thereby significantly exceeding the number of immersed persons in previous approaches~\cite{stotko2019slamcast} without adding or exchanging hardware components.

In summary, the key contributions of this work are
(1) an efficient novel set of filters designed to optimize the performance and scalability of current state-of-the-art telepresence systems at the example of the SLAMCast system~\cite{stotko2019slamcast},
(2) an adaption of the telepresence-specific filters to standalone volumetric 3D reconstruction and
(3) a comprehensive evaluation of the beneficial effect of the proposed set of filters regarding scalability, latency and visual quality.

\section{Related Work}

Telepresence applications for sharing live experiences rely on real-time 3D scene capture.
For this purpose, the underlying scene representation, where the scene is reconstructed based on the fusion of the incoming sensor data, is of particular importance.
Well-established representations include surface modeling in the form of implicit truncated signed distance fields (TSDFs).
Early real-time volumetric reconstruction approaches~\cite{kinectfusion,kinectfusion2} are based on storing the scene model in a uniform grid.
This results in high memory requirements as the data structure is not adapted according to the local presence of a surface.
To improve the scalability to large-scale scenes, further work exploited the sparsity in the TSDF representation, \eg based on moving volume techniques~\cite{Roth:2012,Whelan:2015}, representing scenes in terms of blocks of volumes that follow dominant planes~\cite{Henry:2013} or storing TSDF values only near the actual surface areas~\cite{Chen:2013,niessner,infinitam}.
The individual blocks can be managed using tree structures or hash maps as proposed by Nie{\ss}ner \etal~\cite{niessner} and respective optimizations~\cite{infinitam,Kaehler:2016:Hierarchical,Reichl:2016}.
Furthermore, the replacement of the TSDF representation by a high-resolution binary voxel grid has also been considered by Reichl \etal~\cite{Reichl:2016} to improve the scalability and reduce the memory requirements.
Recent extensions include the detection of loop closures~\cite{Kaehler:2016,Dai:2017,maier2017efficient} to reduce drift artifacts in camera localization as well as multi-client collaborative acquisition and reconstruction of static scenes~\cite{Golodetz2018Collaborative}.

This progress in real-time capturing enabled the development of various telepresence applications.
Early telepresence systems~\cite{kinectfusion,Maimone:2012,Maimone:2012b,Molyneaux:2012,Jones:2014,Fuchs:2014} were designed for room-scale environments and faced the problems of a limited reconstruction quality due to high sensor noise and a reduced resolution.
Relying on an expensive capturing setup with several cameras, GPUs and desktop computers, the Holoportation system~\cite{Orts-Escolano:2016} was designed for high-quality real-time reconstruction of a dynamic room-scale environment based on the Fusion4D system~\cite{Dou:2016} as well as real-time data transmission.
This has been complemented with AR/VR systems to allow immersive end-to-end teleconferencing.
In contrast, interactive telepresence for individual remote users within live-captured static scenes has been addressed by Mossel and Kr{\"o}ter~\cite{mossel} based on voxel block hashing~\cite{niessner,infinitam}.
The limitations of this system regarding high bandwidth requirements, the immersion of only a single remote user into the captured scenarios as well as network interruptions leading to loss of scene parts that are reconstructed in the meantime have been overcome in the recent SLAMCast system~\cite{stotko2019slamcast}.
However, the scalability to immersing large groups of people into on-the-fly captured scenes has not been achieved so far.
In this paper, we directly address this problem by several modifications to the major components involved in telepresence systems.

\section{System Outline}

Akin to previous work, we build our scalable multi-client telepresence system on top of a volumetric scene representation in terms of voxel blocks, \ie blocks of $ 8^3 = 512 $ voxels.
This approach has been well-established by previous investigations in the context of real-time reconstruction~\cite{kinectfusion,kinectfusion2,Whelan:2012,niessner,Chen:2013,Whelan:2015,Newcombe:2015,infinitam,Kaehler:2016:Hierarchical,Kaehler:2016,Dai:2017} and telepresence~\cite{mossel,Orts-Escolano:2016,stotko2019slamcast}.
As shown in \autoref{fig:overview}, current state-of-the-art telepresence systems involving live-captured scenarios rely on the core components of (1) a real-time 3D reconstruction process, (2) a central server process as well as (3) exploration clients.
RGB-D images captured by a single camera are streamed to the reconstruction process that runs on a cloud server and allows on-the-fly camera localization and scene capture via volumetric fusion.
The reconstructed scene data is then passed to the central server process that manages a bandwidth-optimized version of the global model as well as the streaming of these data according to requests by connected exploration clients.
Each exploration client integrates the transmitted scene parts into a locally generated mesh that can be interactively explored with VR devices on their local computers.
In the following, we focus on the extension of such live telepresence systems to the immersion of larger groups of remote users into a live-captured scene at the example of the SLAMCast system~\cite{stotko2019slamcast}.
This requires the optimization of the reconstruction (see \autoref{sec:reconstruction}) and the central server processes (see \autoref{sec:server}).
In contrast, the exploration client receives the compressed and optimized scene representation and is already capable of providing an immersive viewing experience at the remote user's site.

\section{Optimization of the 3D Reconstruction Process}
\label{sec:reconstruction}

\begin{figure*}[t]
    \centering
    \includegraphics[width=\textwidth]{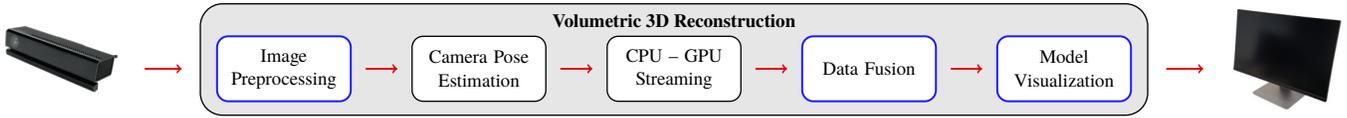}
    \vspace{-3.5mm}
    \caption{General volumetric 3D reconstruction pipeline. Our set of efficient filters designed to improve the performance and scalability of the state-of-the-art live telepresence systems can also be applied to the components of standalone 3D reconstruction (highlighted).}
    \label{fig:reconstruction_pipeline}
\end{figure*}

Since our optimizations are not particularly restricted to the reconstruction process used in the SLAMCast system, we show their application to volumetric 3D reconstruction approaches in general and provide an overview of the respective pipeline (see \autoref{fig:reconstruction_pipeline}).
Here, the surface is represented in terms of implicit truncated signed distance fields (TSDFs) and stored as a sparse unordered set of voxel blocks using spatial hashing~\cite{niessner,infinitam,Kaehler:2016,Dai:2017,Reichl:2016}.
Input to the reconstruction pipeline is an incremental stream of RGB-D images which is processed in an online fashion.
First, the current RGB-D frame is preprocessed where camera-specific distortion effects are removed and a normal map is computed from the depth data.
Afterwards, the current camera pose is estimated either using frame-to-model tracking~\cite{kinectfusion,kinectfusion2,niessner,infinitam,Whelan:2013,Whelan:2015,Reichl:2016} (as also used in the SLAMCast system) or using bundle adjustment for globally-consistent reconstruction~\cite{Kaehler:2016,Dai:2017}.
Using this pose, non-visible voxel block data are streamed out to CPU memory whereas visible blocks in CPU memory are streamed back into GPU memory~\cite{niessner,infinitam,Reichl:2016,maier2017efficient}.
In the next step, new voxel blocks are allocated in the volume and the RGB-D data are fused into the volumetric model.
Finally, a novel view depicting the current state of the reconstruction is generated using raycasting to provide a live feedback to the user during capturing.

\subsection{Image Preprocessing}
\label{sec:preprocessing}

We improve the robustness of the acquired RGB-D data by filtering potentially unreliable data from the depth map.
A further benefit of this operation is the resulting more compact scene model representation.
Inspired by previous work~\cite{Whelan:2015}, we discard samples $ d $ located on stark depth discontinuities by considering the deviations to the depth values $ d_i $ in a $ 7 \times 7 $ neighborhood $ \makeMathBold{N}(d) $.
Due to the limited resolution and the overall noise characteristics of the sensor, such samples are likely to be outliers and might largely deviate from the true depth values.
We extend this filter by further discarding samples $ d $ with a significant amount of missing data in their local neighborhood.
In such regions, which may not only contain depth discontinuities, the depth measurements are also susceptible of being unreliable.
Thus, we consider the set
\begin{equation}
    \makeMathBold{D}_o = \left \{ d \mid \exists i \in \makeMathBold{N}(d) \colon \abs{d - d_i} > c_{d} \lor \abs{\makeMathBold{N}_o(d)} > c_{h} \cdot \abs{\makeMathBold{N}(d)} \right \}
\end{equation}
as outliers where $ c_{d} $ and $ c_{h} $ are user-defined thresholds, $ \makeMathBold{N}(d) $ denotes the neighborhood of the depth sample $ d $ and $ \makeMathBold{N}_o(d) $ the set of neighboring pixels with no valid depth data.
These outliers affect the overall reconstruction quality as well as the model compactness.

\subsection{Data Fusion}
\label{sec:datafusion}

Although potentially unreliable data around stark depth discontinuities have been filtered out during the preprocessing step, there are still samples, \eg around small discontinuities, that do not contribute to the reconstruction and negatively affect the model compactness and streaming performance.
In the voxel block allocation step, these unreliable data unnecessarily enlarge the global truncation region around the unknown surface since all voxel blocks located within the local truncation region around the respective depth samples are considered during allocation.
Traditional approaches tried to remove these blocks afterwards using a garbage collection~\cite{niessner} which requires a costly analysis of the voxel data.
In contrast, we propose a novel implicit filter which reduces the amount of unnecessary block allocations.
By considering only every $ c_{a} $-th pixel per column and row, where $ c_{a} $ is a user-defined control parameter, the depth image is virtually downsampled and the likelihood for an over-sized global truncation region is significantly reduced.
Furthermore, this reduces the number of processed voxels during data fusion which greatly speed-ups the reconstruction and reduces the amount of blocks that are later queued for streaming to the server.
Note that this downsampling is only performed during allocation whereas the whole depth image is still used for data fusion to employ TSDF-based regularization.
In the context of globally-consistent 3D reconstruction using bundle-adjusted submaps~\cite{Kaehler:2016,Golodetz2018Collaborative}, our filter improves the compactness of the respective submap into which the RGB-D data are fused whereas the fusion of the more compact submaps into a single global model would be performed as in previous work.

\subsection{Model Visualization}
\label{sec:isosurface}

In order to provide a decent live preview of the current model state, the generation of such model views should preserve all the relevant scene information while suppressing noise as much as possible.
Furthermore, if frame-to-model tracking is used to estimate the current camera pose, this is also crucial to allow a robust alignment.
We propose a Marching Cubes (MC) voxel block pruning approach which will be described in more detail in \autoref{sec:server}, as it has been carefully designed for the central server process.
Here, we show an adaption of this contribution to standalone volumetric 3D reconstruction where the model is stored implicitly using TSDF voxels.
Each TSDF voxel stores a TSDF value $ D \in [-1; 1] $ and fusion weight $ W \in [0; 255] $ (both compressed using 16-bit linear encoding~\cite{infinitam}) as well as a 24-bit color $ C \in [0; 255]^3 $.
Inspired by the garbage collection approach of point-based reconstruction techniques~\cite{Keller:2013}, we ignore TSDF voxels for raycasting and triangle generation which are currently considered unstable.
These voxels contain only very few, possibly unreliable observations from the input data, so their fusion weight falls below a user-defined threshold $ c_{w} $:
\begin{equation}
    \makeMathBold{V}_o^{TSDF} = \left \{ (D, W, C) \mid W < c_{w} \right \}
\end{equation}
However, in contrast to previous garbage collection approaches~\cite{niessner,Keller:2013}, we do not remove these voxel blocks but only ignore them.
This avoids accidental removal of blocks that might become stable at a future time when this scene part is also partially stored in a different submap or revisited by the user or another client in multi-client acquisition setups~\cite{Kaehler:2016,Golodetz2018Collaborative}.
Furthermore, by ignoring unstable data, the raycasted view will also be consistent with the exploration client's version of the 3D model.

\section{Optimization of the Central Server Process}
\label{sec:server}

\begin{algorithm}[t]
    \footnotesize
    \caption{Our optimized server voxel block data integration}
    \label{alg:server}
    \begin{algorithmic}[1]
        \Require Received TSDF voxel block positions $ \makeMathBold{P}^{TSDF} $ and voxel data $ \makeMathBold{V}^{TSDF} $
        \Ensure Voxel block position list $ \makeMathBold{P}^{MC} $ for updating the stream sets
        \State $ \makeMathBold{M}^{TSDF} \leftarrow \mathtt{allocateBlocks}(\makeMathBold{P}^{TSDF}) $
        \State $ \makeMathBold{M}^{TSDF} \leftarrow \mathtt{copyVoxelData}(\makeMathBold{V}^{TSDF}) $
        \State $ \makeMathBold{P}^{MC} \leftarrow \mathtt{createBlockUpdateSet}(\makeMathBold{P}^{TSDF}) $
        \State $ \makeMathBold{V}^{MC}, \makeMathBold{F}^{MC} \leftarrow \mathtt{computeVoxelData}(\makeMathBold{P}^{MC}, \makeMathBold{M}^{TSDF}) $
        \State $ \makeMathBold{M}^{MC} \leftarrow \mathtt{allocateNonEmptyBlocks}(\makeMathBold{P}^{MC}, \makeMathBold{F}^{MC}) $
        \State $ \makeMathBold{M}^{MC} \leftarrow \mathtt{copyNonEmptyVoxelData}(\makeMathBold{V}^{MC}, \makeMathBold{F}^{MC}) $
        \State $ \makeMathBold{M}^{MC} \leftarrow \mathtt{pruneEmptyBlocks}(\makeMathBold{P}^{MC}, \makeMathBold{F}^{MC}) $
        \State $ \makeMathBold{P}^{MC} \leftarrow \mathtt{pruneBlockUpdateSet}(\makeMathBold{P}^{MC}, \makeMathBold{P}_A^{MC}, \makeMathBold{F}^{MC}) $
        \State $ \makeMathBold{P}_A^{MC} \leftarrow \mathtt{updateNonEmptyBlockSet}(\makeMathBold{P}_A^{MC}, \makeMathBold{P}^{MC}) $
    \end{algorithmic}
\end{algorithm}

Beyond optimizations in the reconstruction process, the scalability of a live telepresence system also relies on the optimization of its central server process that takes care of managing the reconstructed global scene model as well as the stream states and requests by connected exploration clients.
In this regard, we show respective optimizations at the example of the recently published SLAMCast system~\cite{stotko2019slamcast}.
In comparison to the standard voxel block data integration at the server side, we propose a further filtering step which discards empty or unstable voxel blocks that contain only very few or none observations from the input RGB-D image data.
This significantly improves the streaming performance and scalability and allows the immersion of groups of people.
The individual steps of our optimized integration approach are shown in \autoref{alg:server}.

Similar to the original SLAMCast system, we first integrate the TSDF voxel block positions $ \makeMathBold{P}^{TSDF} $ and voxel data $ \makeMathBold{V}^{TSDF} $ into the global TSDF voxel block model $ \makeMathBold{M}^{TSDF} $ of the central server process.
Afterwards, we update the global MC voxel block model $ \makeMathBold{M}^{MC} $ which is optimized for streaming and stores a Marching Cubes index $ I \in [0; 255] $ as well as a 24-bit color $ C \in [0; 255]^3 $ in each MC voxel.
For this purpose, we create the set $ \makeMathBold{P}^{MC} $ of MC voxel block positions requiring an update as well as a set of flags $ \makeMathBold{F}^{MC} $ and the respective MC voxel data $ \makeMathBold{V}^{MC} $ by performing the Marching Cubes algorithm on the corresponding TSDF voxels~\cite{Lorensen:1987:MCH}.
The flags $ \makeMathBold{F}^{MC} $ indicate whether a block will generate reliable triangles and are constructed by analyzing the Marching Cubes indices $ I $ of the MC voxels as well as the fusion weight $ W $ of the corresponding TSDF voxels.
Therefore, the following set $ \makeMathBold{V}_o^{MC} $ of voxels either does not contain surface information in terms of triangles or would generate unstable triangle data:
\begin{equation}
    \makeMathBold{V}_o^{MC} = \left \{ (I, C) \mid I = 0 \lor I = 255 \lor W < c_{w} \right \}
\end{equation}
We only allocate those blocks in the MC voxel block model $ \makeMathBold{M}^{MC} $ that are flagged and prune blocks that are currently not flagged.
This minimizes the amount of scene data that are streamed to the exploration clients.
Finally, we integrate the generated MC voxel data $ \makeMathBold{V}^{MC} $ of the flagged blocks.
We do not prune the TSDF voxel block model $ \makeMathBold{M}^{TSDF} $ which would otherwise lead to potential artifacts, \ie missing geometry at block boundaries, since currently empty blocks might be needed for future updates.

In contrast to the MC voxel block model, pruning the list of updated MC voxel block positions $ \makeMathBold{P}^{MC} $ in the same way would introduce artifacts at the exploration client side since they may already have received a previous version of blocks that have been pruned in the meantime.
To properly handle updates, we manage the update set $ \makeMathBold{P}_A^{MC} $ containing all voxel block positions that were considered for streaming in the past.
We generate the list of updated MC voxel blocks by only considering the ones which either generated triangles in the past or with the current update.
Finally, after the MC voxel blocks have been integrated into the volume and the list of updated block positions has been generated, we update the set $ \makeMathBold{P}_A^{MC} $ by inserting all currently integrated voxel block positions.

\section{Evaluation}
\label{sec:evaluation}

We tested our highly scalable telepresence system on a variety of different datasets and analyzed several aspects such as system scalability, streaming latency and visual quality.
For a quantitative comparison of the proposed contributions, we considered the following variants of our system:
\begin{itemize}[leftmargin=1em]\setlength\itemsep{0.0em}
    \item \textbf{Base (B)}:
        Our 3D reconstruction and streaming system with deactivated filtering contributions, yielding equivalent performance to SLAMCast~\cite{stotko2019slamcast}.
    \item \textbf{Base + Depth Discontinuity Filter (B+DDF)}:
        The base approach with an additional depth map filtering at discontinuities with $ c_{h} $ = 0.25, $ c_{d} $ = 0.2m (see \autoref{sec:preprocessing}).
    \item \textbf{Base + Voxel Block Allocation Downsampling (B+VBAD)}:
        The base approach with an additional virtual downsampling at the voxel block allocation stage with $ c_{a} $ = 4 (see \autoref{sec:datafusion}).
    \item \textbf{Base + MC Voxel Block Pruning (B+MCVBP)}:
        The base approach with an additional pruning of empty MC voxel blocks at the server side with $ c_{w} $ = 2.0 (see \autoref{sec:isosurface} and \autoref{sec:server}).
    \item \textbf{Ours}:
        Our approach incorporating all filtering contributions.
\end{itemize}
The filter sizes and thresholds as described above were determined empirically using several datasets.
For validation, we used different real-world datasets recorded with an ASUS Xtion Pro (\emph{lounge}, \emph{copyroom})~\cite{zhou2013dense} and a Kinect v2 (\emph{heating\_room}, \emph{pool})~\cite{stotko2019slamcast} as well as synthetic data (\emph{lr kt2} with simulated noise)~\cite{handa2014benchmark}.
Throughout the experiments, we used three computers where each of them takes the role of one part of the telepresence system, \ie 3D reconstruction process (RC), central server process (S) and exploration client (EC).
All computers were equipped with an Intel Core i7-4930K CPU and 32GB RAM and a NVIDIA GTX 1080 GPU with 8GB VRAM and connected via a local network.
We replaced the exploration client by a benchmark client which starts requesting voxel blocks with a fixed frame rate of 100Hz when the reconstruction process starts.
Furthermore, the reconstruction process uses a fixed reconstruction speed of 30Hz matching the framerate of the used datasets.
We set the voxel size to 5mm as well as the truncation region to 60mm and used hash map/set sizes of $ 2^{20} $ and $ 2^{22} $ buckets as well as GPU and CPU voxel block pool sizes of $ 2^{19} $ and $ 2^{20} $ blocks, thereby following previous work~\cite{stotko2019slamcast}.

\subsection{System Scalability}
\label{sec:scalability}

\begin{table}[t]
    \fontsize{7pt}{8pt}\selectfont
    \centering
    \caption{Maximum number of exploration clients (ECs) that the server can handle without any delay compared to a single client. Instead of using different package sizes with a fixed request rate of 100Hz, we use a fixed size of 512 and vary the rate accordingly to demonstrate the highest possible scalability. If empty MC voxel block pruning is used (B+MCVBP and Ours), the sizes of the TSDF and MC voxel block models differ and we list both ($ \makeMathBold{M}^{MC} / \makeMathBold{P}_A^{MC} (\makeMathBold{M}^{TSDF}) $).}
    \begin{tabular}{llC{0.025\textwidth}C{0.06\textwidth}C{0.1275\textwidth}}
        \toprule
        Approach & Dataset              & Max. ECs & Request Rate [Hz] & Model Size [\# $ \times 10^3 $ \par MC Voxel Blocks] \\
        \midrule
        \multirow{5}{*}{B}
                 & \emph{lounge}        &  5       & 100               & 314 \\
                 & \emph{copyroom}      &  9       &  50               & 228 \\
                 & \emph{heating\_room} &  3       & 100               & 850 \\
                 & \emph{pool}          &  5       & 100               & 590 \\
                 & \emph{lr kt2}        &  1       & 200               & 834 \\
        \midrule
        \multirow{5}{*}{B+DDF}
                 & \emph{lounge}        &  9       &  50               & 270 \\
                 & \emph{copyroom}      & 10       &  50               & 230 \\
                 & \emph{heating\_room} &  9       &  50               & 443 \\
                 & \emph{pool}          & 10       &  50               & 379 \\
                 & \emph{lr kt2}        & 10       &  50               & 227 \\
        \midrule
        \multirow{5}{*}{B+VBAD}
                 & \emph{lounge}        &  8       &  50               & 264 \\
                 & \emph{copyroom}      &  5       & 100               & 226 \\
                 & \emph{heating\_room} &  4       & 100               & 622 \\
                 & \emph{pool}          &  8       &  50               & 446 \\
                 & \emph{lr kt2}        &  1       & 200               & 550 \\
        \midrule
        \multirow{5}{*}{B+MCVBP}
                 & \emph{lounge}        & 21       &  12               &  47 /  51 (314) \\
                 & \emph{copyroom}      &  9       &  50               &  65 /  75 (298) \\
                 & \emph{heating\_room} &  8       &  25               & 120 / 127 (850) \\
                 & \emph{pool}          & 13       &  25               & 104 / 108 (590) \\
                 & \emph{lr kt2}        &  7       &  25               &  64 /  64 (834) \\
        \midrule
        \multirow{5}{*}{Ours}
                 & \emph{lounge}        & 25       &  12               &  44 /  47 (240) \\
                 & \emph{copyroom}      & 18       &  25               &  57 /  67 (202) \\
                 & \emph{heating\_room} & 27       &  12               &  90 /  94 (352) \\
                 & \emph{pool}          & 28       &  12               &  95 /  99 (317) \\
                 & \emph{lr kt2}        & 26       &  12               &  53 /  55 (201) \\
        \bottomrule
    \end{tabular}
    \label{tab:scalability}
\end{table}

\begin{figure*}[t]
    \centering
    \includegraphics[width = \textwidth, height = 0.5\textheight, keepaspectratio]{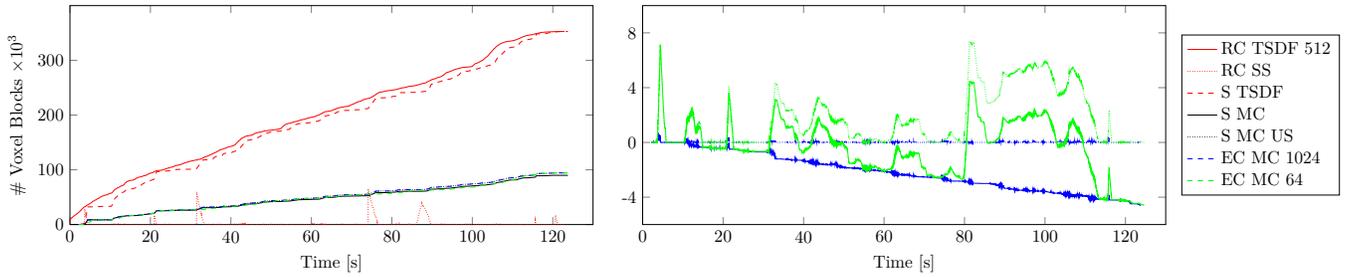}
    \vspace{-3.5mm}
    \caption{Streaming progress and latency between server (S) and exploration client (EC) over time for the \emph{heating\_room} dataset using our full system. Left: Absolute model sizes for the highest and lowest chosen package size. Right: Relative size differences between S and EC (w.r.t. model size S MC and update set size S MC US).}
    \label{fig:completeness_over_time:heating_room}
    \vspace{4mm}
\end{figure*}

\begin{figure*}[t]
    \centering
    \includegraphics[width = \textwidth, height = 0.5\textheight, keepaspectratio]{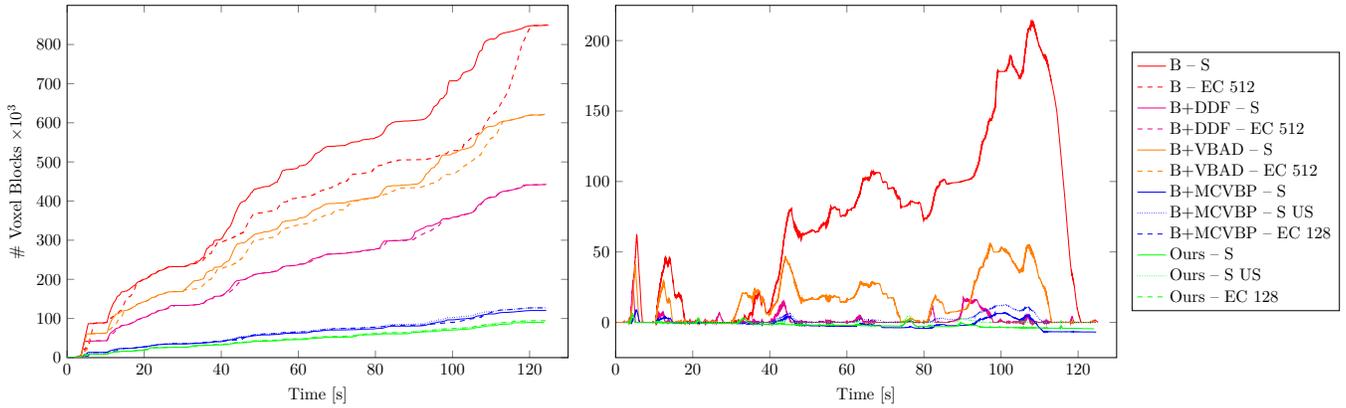}
    \vspace{-3.5mm}
    \caption{Streaming progress and latency between server (S) and exploration client (EC) over time for the \emph{heating\_room} dataset for each system variant. Left: Absolute model sizes. Right: Relative size differences between S and EC.}
    \label{fig:completeness_over_time_approaches:heating_room}
    \vspace{2mm}
\end{figure*}

In this section, we will evaluate the scalability of our system in comparison to the baseline SLAMCast approach (see \autoref{tab:scalability}).
In contrast to the following evaluations, the benchmark client discards the received data which allows for running all benchmark clients on a single computer without an overhead.
Furthermore, rather than lowering the package size, we used a fixed package size of 512 voxel blocks and lower the request rate accordingly.
This significantly reduced the constant overheads of kernel calls and memory copies and introduces only a minimal delay in the range of milliseconds which made it the preferred setting for handling a large number of clients.
For an appropriate choice of the streaming rate, we determined the lowest package size which still allows the benchmark client to retrieve the whole model with an acceptable delay of at most one second (see supplemental material for a detailed analysis).
Then, we measured the maximum number of benchmark clients that the server could handle without introducing a further delay.
While the original SLAMCast system was only able to handle around 3-5 clients in general, both filters at the reconstruction side (B+DDF and B+VBAD) raised this limit to up to 10 clients.
Since there is a tracking loss at the end of the \emph{copyroom} sequence resulting in a slightly higher delay, a higher request rate was chosen and the scalability decreased accordingly.
Although the number of MC voxel blocks is significantly lower after pruning (B+MCVBP), we observed that the general performance is similar to the depth discontinuity filter approach.
Here, the TSDF voxel block model has the same size as in the base approach and is, hence, considerably larger than in the other approaches.
In contrast, our full system reduces the request rate requirements to 12Hz for most scenes making it the preferred choice for this parameter.
This significantly improves its scalability to more than 24 clients in all scenes which is sufficient for applications in education, entertainment or collaboration scenarios.

\subsection{Latency and Streaming Progress Analysis}
\label{sec:latency}

\begin{figure*}[t]
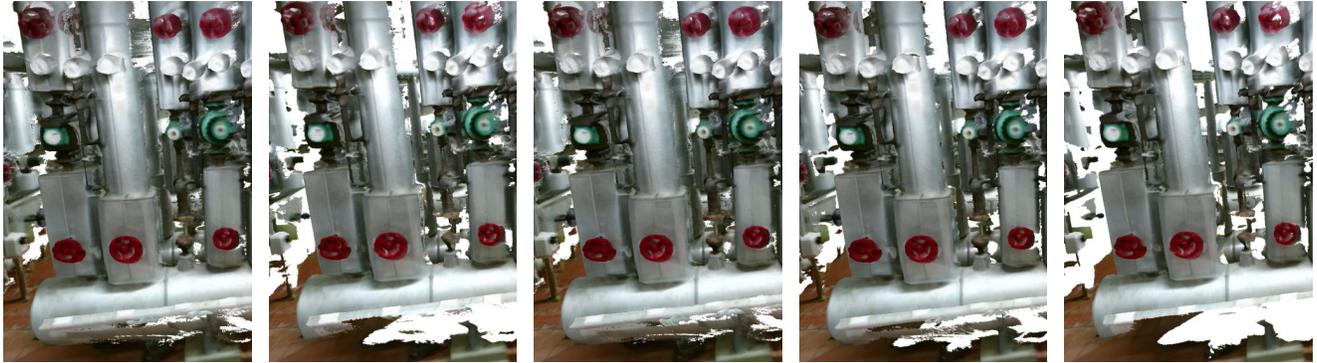
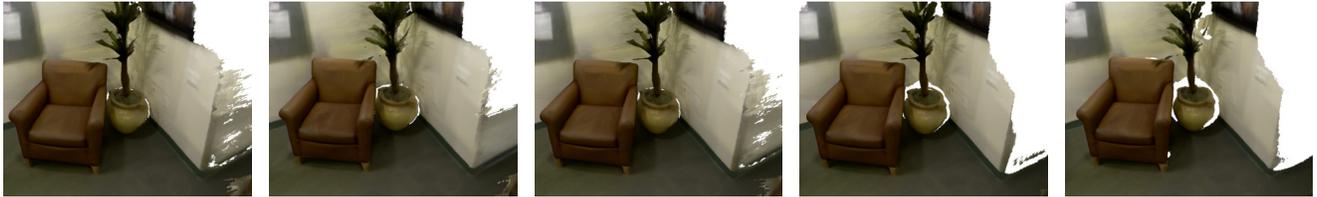

    \centering
    \subfigure[\emph{heating\_room}: B, \newline 16.5ms (8.6ms), 3482MB]
    {
        \centering
        \includegraphics[width = 0.185\textwidth, height = 0.25\textheight, keepaspectratio]{figures/heating_room_B.png}
    }
    \subfigure[\emph{heating\_room}: B+DDF, \newline 10.3ms (4.4ms), 1815MB]
    {
        \centering
        \includegraphics[width = 0.185\textwidth, height = 0.25\textheight, keepaspectratio]{figures/heating_room_B+DDF.png}
    }
    \subfigure[\emph{heating\_room}: B+VBAD, \newline 13.2ms (6.6ms), 2548MB]
    {
        \centering
        \includegraphics[width = 0.185\textwidth, height = 0.25\textheight, keepaspectratio]{figures/heating_room_B+VBAD.png}
    }
    \subfigure[\emph{heating\_room}: B+MCVBP, \newline 16.1ms (8.7ms), 3482MB]
    {
        \centering
        \includegraphics[width = 0.185\textwidth, height = 0.25\textheight, keepaspectratio]{figures/heating_room_B+MCVBP.png}
    }
    \subfigure[\emph{heating\_room}: Ours, \newline 9.7ms (3.3ms), 1442MB]
    {
        \centering
        \includegraphics[width = 0.185\textwidth, height = 0.25\textheight, keepaspectratio]{figures/heating_room_Ours.png}
    }
    \\\hspace{0.1mm}
    \subfigure[\emph{lounge}: B, \newline 10.9ms (5.0ms), 1286MB]
    {
        \centering
        \includegraphics[width = 0.185\textwidth, height = 0.25\textheight, keepaspectratio]{figures/lounge_B.png}
    }
    \subfigure[\emph{lounge}: B+DDF, \newline 10.0ms (4.4ms), 1106MB]
    {
        \centering
        \includegraphics[width = 0.185\textwidth, height = 0.25\textheight, keepaspectratio]{figures/lounge_B+DDF.png}
    }
    \subfigure[\emph{lounge}: B+VBAD, \newline 10.0ms (5.4ms), 1081MB]
    {
        \centering
        \includegraphics[width = 0.185\textwidth, height = 0.25\textheight, keepaspectratio]{figures/lounge_B+VBAD.png}
    }
    \subfigure[\emph{lounge}: B+MCVBP, \newline 10.9ms (5.2ms), 1286MB]
    {
        \centering
        \includegraphics[width = 0.185\textwidth, height = 0.25\textheight, keepaspectratio]{figures/lounge_B+MCVBP.png}
    }
    \subfigure[\emph{lounge}: Ours, \newline 9.7ms (4.3ms), 983MB]
    {
        \centering
        \includegraphics[width = 0.185\textwidth, height = 0.25\textheight, keepaspectratio]{figures/lounge_Ours.png}
    }
    \caption{Comparison of visual quality, mean runtime (and standard deviation) as well as memory requirements for each system variant. All individual contributions reduced the amount of reconstruction artifacts while improving the overall reconstruction performance.}
    \label{fig:visual_comparison}
\end{figure*}

In addition to the scalability analysis, we also measured the streaming latency over time (see \autoref{fig:completeness_over_time:heating_room}).
Similar to the original SLAMCast approach, our system has a small delay between the reconstruction process and the server process due to the shared streaming strategy.
However, our optimized server model prunes unreliable or irrelevant blocks which results in a very low latency between the server and the exploration client.
We also compared the latency between the largest and smallest chosen package size, \ie 1024 and 64 blocks/request.
Here, the model size of the exploration client is close to the size of the server's update set $ \makeMathBold{P}_A^{MC} $ indicating a very fast and low-latent streaming while the gap to the minimal size of the server model $ \makeMathBold{M}^{MC} $ increases over time.
Note that these two sizes are the bounds for the exploration client's model size and clients which have reconnected, \eg due to network outages, will receive a slightly more compact model closer to the lower bound.
Reducing the package size from 1024 to 64 blocks significantly reduces the bandwidth requirements (see supplemental material for a detailed analysis) and leads to a slightly worse latency when the reconstruction process queues the currently visible voxel blocks for streaming.

In \autoref{fig:completeness_over_time_approaches:heating_room}, we also compared the different system variants regarding streaming progress and latency.
For a fair comparison between the approaches, the package size is chosen such that the mean bandwidths are similar, \ie around 15Mbit/s.
Here, we also considered the size of the update set $ \makeMathBold{P}_A^{MC} $ in addition to size of the server model $ \makeMathBold{M}^{MC} $ when empty MC voxel block pruning is enabled (B+MCVBP and Ours).
In these scenarios, the number of voxel blocks transmitted to the exploration client bound by these two sizes is typically close to the upper bound $ \makeMathBold{P}_A^{MC} $.
In comparison to the baseline, both filtering approaches at the reconstruction side (B+DDF and B+VBAD) reduce the latency significantly.
Similar results can be seen when empty MC voxel blocks are pruned (B+MCVBP).
Whereas all of these approaches still introduce a noticeable delay at the time steps 40s and 90-100s, our full system is capable of streaming the reconstructed model with almost no delay across the whole sequence.
Additional results regarding bandwidth and streaming latency over time are provided in the supplemental material.

\subsection{Visual Quality}

In order to demonstrate the benefit for standalone volumetric 3D reconstruction, we also provide a qualitative comparison regarding the visual quality of the reconstructed 3D models as well as the respective runtime and memory requirements for the individual system variants (see \autoref{fig:visual_comparison}).
In general, all approaches generated detailed and accurate 3D models from the noisy RGB-D input data.
However, without filtering, there might be some artifacts around depth discontinuities as well as in regions which have not been fully observed by the camera.
These artifacts affect the overall visual experience and lead to high runtime and memory requirements.
Using virtual downsampling at the voxel block allocation stage (B+VBAD), we obtain almost identical 3D models but the computational burden is significantly lower since the number of empty blocks within the model is reduced.
In contrast, filtering depth samples at depth discontinuities (B+DDF) or unreliable triangle data during Marching Cubes (B+MCVBP) reduces the amount of artifacts in the aforementioned regions.
Note that in standalone 3D reconstruction, voxel block pruning (B+MCVBP) mainly affects the triangulation step at the end of the capturing session which leads to results similar to the base approach regarding runtime and memory.
Our full system enhances the visual quality even further and almost completely removes artifacts without sacrificing the overall model completeness.
Here, we observe improvements of 10-40\% and 25-60\% for the runtime and memory footprint respectively depending on the scene.
The objects in the \emph{lounge} scene have been captured at a much smaller distance and from more angles than in the \emph{heating\_room} scene which leads to less unreliable input data and, hence, a lower impact of our outlier filtering approach.
Additional performance measurements and results are provided in the supplemental material.
In the context of live remote collaboration, a slightly less complete model can be beneficial and helps to identify regions that still need to be captured and reliably reconstructed.
This, in turn, might even increase the model completeness and accuracy since the scene is more thoroughly acquired by the user.

\subsection{Limitations}

Despite the significant improvements in terms of scalability, latency and visual quality, our system still has some limitations.
Since our work is based on the SLAMCast system, misalignments within the reconstruction might occur due to fast camera movement.
While this problem has been addressed by loop-closure techniques~\cite{Dai:2017,Kaehler:2016}, their integration into live telepresence systems is still highly challenging.
Furthermore, too aggressive virtual downsampling during voxel block allocation might lead to holes in the final model when some blocks covering distant objects are always skipped and, hence, never allocated.
However, this is only problematic for long-range devices whereas typical RGB-D cameras have a smaller range of up to 5 meter which is still sufficient for most scenarios.

\section{Conclusion}

We presented a highly scalable multi-client live telepresence system which allows immersing a large number of people into a live-captured environment.
For this purpose, we used well-established systems and proposed several optimizations regarding scalability, latency, and visual quality.
While our contributions are designed with the telepresence system in mind, we also show their application to standalone volumetric 3D reconstruction approaches.
As demonstrated in a comprehensive evaluation, our novel system allows the immersion of more than 24 people within the same scene using consumer hardware.

\acknowledgments{
This work was supported by the DFG projects KL 1142/11-1 (DFG Research Unit FOR 2535 Anticipating Human Behavior) and KL 1142/9-2 (DFG Research Unit FOR 1505 Mapping on Demand).
}

\bibliographystyle{abbrv-doi}

\bibliography{literature}

\end{document}